\documentclass[12pt]{article}
\usepackage{amsmath}
\usepackage{amsfonts}
\usepackage{amssymb}
\usepackage{graphics}
\usepackage{graphicx}
\usepackage{amssymb}
\usepackage{cite}
\usepackage{setspace}
\usepackage{appendix}
\usepackage{graphicx} 

\usepackage{booktabs} 
\usepackage{array} 
\usepackage{paralist} 
\usepackage{verbatim} 
\usepackage{subfig} 

\usepackage{fancyhdr} 
\usepackage{sectsty}
\allsectionsfont{\sffamily\mdseries\upshape} 

\usepackage[nottoc,notlof,notlot]{tocbibind} 

\newcommand{\inv}{^{-1}}                                               
\newcommand{\beq}{\begin{equation}}
\newcommand{\eeq}{\end{equation}}
\newcommand{\bs}{\boldsymbol}
\numberwithin{equation}{section}

\title{Quantum Mechanics in Noninertial Reference Frames:  Violations
of the Nonrelativistic Equivalence Principle}
\author{W. H.~Klink$^1$ and S.~Wickramasekara$^{1,2}$\\ 
$^1$ Department of Physics and Astronomy\\University of
Iowa\\
 Iowa City, IA 52242\\
$^2$ Department of Physics\\
Grinnell College\\ 
Grinnell, IA 50112}
\date{}
\begin{document}
\maketitle
\begin{abstract}
In previous work we have developed a formulation of quantum mechanics in non-inertial reference frames. This formulation is grounded in a class of unitary cocycle representations of what we have called the Galilean line group, the generalization of the Galilei group that includes transformations amongst non-inertial reference frames. These representations show that in quantum mechanics, just as is the case in classical mechanics,  the transformations to accelerating reference frames give rise to fictitious forces. A special feature of these previously constructed representations is that they all respect the non-relativistic equivalence principle,  wherein the fictitious forces associated with linear acceleration can equivalently be described by gravitational forces.  In this paper we exhibit a large class of cocycle  representations of the Galilean line group that violate the equivalence principle.  Nevertheless the classical mechanics analogue of these cocycle representations all respect the equivalence principle.
\end{abstract}
\section{Introduction}
This paper is a sequel to two previous papers \cite{bk1,sw1} on the formulation of Galilean quantum mechanics in non-inertial 
reference frames and its implications for the equivalence principle and gravity. Both of these papers were grounded on the same philosophical principle, namely 
that the structure of quantum physics is fundamentally determined by the unitary representations of the symmetry group that implements the principle of 
relativity for the relevant spacetime. Thus, Galilean quantum mechanics is grounded on the unitary projective representations of the Galilei group while 
Lorentzian quantum mechanics, including quantum field theory,  is grounded on the unitary representations of the Poincar\'e group. Quantum mechanics and 
quantum field theory on a deSitter spacetime may also be viewed as being grounded on the unitary representations of the deSitter group. 
Both Galilei and Poincar\'e groups involve transformations amongst only inertial reference frames, the privileged class of reference frames with respect to 
which accelerations of bodies can be understood as the result of forces impressed upon them.  Standard quantum mechanics, too, being grounded on the representations 
of either the Galilei group or the Poincar\'e group, is consequently meaningful as a physical theory with respect to this class of privileged reference frames. 
The primary goal of both \cite{bk1} and \cite{sw1} was to consider the symmetry group of transformations amongst non-inertial reference frames and 
extend the view that quantum physics is grounded on the principle of relativity to include non-inertial reference frames. In such a setting, a natural question that arises 
is the role of the equivalence principle and fictitious forces  in quantum mechanics. As is well known, 
Einstein initially used accelerated systems and their accompanying fictitious forces as a guide to understanding the nature of gravitation in classical physics.  Whether this also can be done for quantum theory will be investigated in this and subsequent papers. 

Both \cite{bk1} and \cite{sw1} introduce the concept of the line group of the three dimensional Euclidean group, the group of (analytic) functions on 
the real line taking values on the Euclidean group, as the group of acceleration transformations.  The main difference between the two papers is that in \cite{sw1} 
time translations are included in the transformation group leading to a semidirect product of the Euclidean line group and the real line, called the Galilean line group, 
while in \cite{bk1} time translations 
are not introduced explicitly and the analysis is carried out by way of transformation properties of Schr\"odinger's equation under the Euclidean line group.  As a result, 
the representations of the two studies are different, particularly with regard to their cocycle structure. However, when the
time translation parameter is set to zero, the representations of \cite{sw1} reduce to those of \cite{bk1} up to a coboundary (i.e., a removable phase factor). 
Both studies come to similar conclusions:
\begin{enumerate}
\item  A unitary irreducible cocycle representation of acceleration transformations can be constructed on the same Hilbert space that carries 
 a given unitary irreducible projective representation of the Galilei group. 
 \item The Hamiltonian for such a representation, when compared to the Hamiltonian that follows from the Galilei group, 
 has an additive term that  depends on the acceleration.  Moreover, this term is proportional to the inertial mass, just as 
one would expect from the classical equivalence principle. 
\item These properties lead to the natural interpretation of the term of the Hamiltonian that arises from the non-inertial nature of a reference frame as a ``fictitious potential 
energy" term. To the extent the equality of inertial and gravitational masses holds, this fictitious potential energy term in turn has interpretation as gravitational potential energy, 
obtained now as a fundamental quantum mechanical entity (as opposed to, say, ``quantizing" a pre-exisiting classical field). 
\item Gravitational fields obtainable this way have zero spatial gradient (potential energy linear in position), though they may have arbitrary time dependence. This is the main limitation of \cite{bk1,sw1}, although 
it is not surprising that acceleration transformations that are functions of time alone give rise to gravitational fields that are functions of time alone. 
\end{enumerate}
These results, particularly the property that the fictitious potential energy term is proportional to the inertial mass, depend on the choices of the two-cocycle made in constructing the representations of 
\cite{bk1} and \cite{sw1}. If we adopt the criterion that any physically meaningful representation of acceleration transformations must contain a physically meaningful representation of the Galilei 
group as a subrepresentation, then we must naturally consider cocycle representations, rather than true vector representations,  of acceleration transformations since it is only such representations of the Galilei group that are quantum 
mechanically meaningful. An acceptable two-cocycle of acceleration transformations must have the property that it reduce to a two-cocycle of the Galilei group when acceleration transformations are  
restricted to Galilean transformations. While they fulfill this property, the particular choices of two-cocycles made in \cite{bk1,sw1} are not the only ones to do so. The main purpose of this paper 
to present a study of the cohomology of the Galilean line group. In particular, we will show that there exist two-cocycles of the Galilean line group
that, while having the correct reduction property with respect to the Galilei group, lead to violations of the equivalence principle. 
Nevertheless, in the classical limit, the equivalence principle is restored for all of these general cocycle representations. 

The structure of the paper is as follows. In section \ref{sec2}, we introduce the Galilean line group and present a short summary of the results of \cite{bk1,sw1}. In section \ref{sec4}, we will then construct a more general class of possible extensions of the Galilean line group and construct their corresponding unitary representations. We will show that  these representations do not uphold the equivalence principle.  This leads to the question whether it is also possible to violate the
equivalence principle at the classical level.  We investigate this
possibility in section \ref{sec5} by constructing a class of generating functions for classical
systems which all generate the same acceleration transformations and lead to equations of motion
satisfying the equivalence principle. 
Finally, in section \ref{sec6} we summarize our results and provide a conclusion. In appendix \ref{sec3}, we give a short  review of 
 the cohomology groups of a group and how the second cohomology group determines inequivalent extensions of the group. This review provides 
 the mathematical framework for the extensions of the Galilean line group considered in section \ref{sec4}. 

\section{Review of Galilean line group and its representations}\label{sec2}
We consider transformations amongst non-inertial reference frames in a Galilean spacetime. Recall that Galilean 
transformations between inertial reference frames are defined by 
\begin{eqnarray}
\bs{x}'&=&R\bs{x}+\bs{v}t+\bs{a}\nonumber\\
t'&=&t+b\label{2.1}
\end{eqnarray}
where $R$, $\bs{v}$, $\bs{a}$  and $b$ are, respectively, rotations, velocity boosts, space translations and time translations. Note that, unlike Lorentz boosts, 
Galilean boosts do not affect the time coordinate of the manifold. Their action on the space coordinates, $\bs{v}t$, is simply a translation that depends on time linearly. 
Therefore, we can write  \eqref{2.1} more compactly as 
\begin{eqnarray}
\bs{x}'&=&R\bs{x}+\bs{a}(t)\nonumber\\
t'&=&t+b\label{2.2}
\end{eqnarray}
where $\bs{a}(t)=\bs{a}+\bs{v}t$. This structure suggests a generalization of $\bs{a}(t)$ to include space translations with arbitrary time dependence as a way of transforming between accelerating reference frames, such as 
$\bs{a}(t)=\frac{1}{2}\bs{a}t^2$. Likewise, we can accommodate rotational accelerations by letting the rotation matrices $R$ be functions of time. Therewith, we have the transformations 
\begin{eqnarray}
\bs{x}'&=&R(t)\bs{x}+\bs{a}(t)\nonumber\\
t'&=&t+b\label{2.3}
\end{eqnarray}
For the sake of the simplicity of notation, let us denote the transformations \eqref{2.3} simply by $\{(R,\bs{a},b)\}$ with the understanding that $R$ and $\bs{a}$ are now functions of time 
that get evaluated at the time coordinate of the spacetime point on which they act.  By the successive operation of two elements on a given spacetime point $(\bs{x},t)$, we obtain 
a composition rule for the set $\{(R,\bs{a},b)\}$:
\begin{equation}
(R_2, \bs{a}_2, b_2)(R_1, \bs{a}_1, b_1) = ((\Lambda_{b_1} R_2) R_1 , \ (\Lambda_{b_1} R_2)\bs{a}_1 + \Lambda_{b_1} \bs{a}_2 ,\  b_1 + b_2)\label{2.5}.
\end{equation}
where $\Lambda_b$ is the shift operator $(\Lambda_bf)(t)=f(t+b)$ and the products of time dependent $R$ and $\bs{a}$ stand for point-wise multiplication, e.g., 
$(\Lambda_{b_1}R_2)R_1(t):=(\Lambda_{b_1}R_2)(t)R_1(t)=R_2(t+b_1)R_1(t)$. 

The associativity of the composition  rule \eqref{2.5} follows from that fact that $b\to \Lambda_b$ is a homomorphism. 
Also, each element $(R,\bs{a},b)$ has an inverse under \eqref{2.5}:
\begin{equation}
(R, \bs{a}, b)\inv = (\Lambda_{-b} R\inv, -\Lambda _{-b}(R\inv \bs{a}) , -b).\label{2.6}
\end{equation}
Therefore, the set of time dependent rotations and space translations, along with time translations, form a group under \eqref{2.5}. In \cite{sw1}, this group was called the Galilean line group. 
We denote it here by ${\cal G}$. 

Note that time dependent rotations ${\cal R}:=\{(R,\bs{0},0)\}$,  time dependent spatial translations ${\cal A}:=\{(I,\bs{a},0)\}$ and time translations ${\cal B}:=\{(I,0,b)\}$ are all 
subgroups of the Galilean line group.  The subgroup ${\cal E}(3):=\{(R,\bs{a},0)\}$, which is 
the group of maps from the real line to the Euclidean group in three dimensions, was called the Euclidean line group in \cite{bk1}. 
It describes all transformations between linearly and rotationally accelerating reference frames. The Euclidean line group is the semidirect product of 
${\cal A}$ and ${\cal R}$: 
\begin{equation}
\mathcal{E}(3)=\mathcal{A}\rtimes\mathcal{R}.\label{2.7}
\end{equation}
Furthermore, from \eqref{2.5} we note that each $(I,\bs{0},b)\in{\cal B}$ induces an automorphism on the subgroup ${\cal E}(3)$, given by $\Lambda_b$. 
Therefore,  
\begin{equation}
\mathcal{G}  = {\mathcal E}(3)\rtimes_{\Lambda}\mathcal{B}= ( \mathcal{A} \rtimes \mathcal{R} ) \rtimes  _{\Lambda} \mathcal{B}.\label{2.8}
\end{equation}
That is, the Galilean line group is the semidirect product of the time translation subgroup and the line group of the three dimensional Euclidean group\footnote{\label{footnote1}{Note 
that our use of the external semidirect product differs slightly from the canonical definition.
 Looking at the subgroup of space and time translations, we have $(a_2, b_2)(a_1, b_1) = (\Lambda_{b_1}a_2 + a_1, b_2 + b_1)$, whereas a connonical external semi-direct product would 
 give $(a_2, b_2)(a_1, b_1) = (a_2 + \Lambda_{b_2}a_1, b_2 + b_1)$.     
Had we chosen our notation so that multiple transformations composed from left to right instead of right to left, then the stucture of the Galilei Galilean Line Group would look exactly like that of a semidirect product. That is to say, our semidirect product definition is the dual of the more standard one.}}.

The point of view advanced in \cite{bk1,sw1} is that quantum mechanics in non-inertial reference frames can be grounded in unitary representations of this group. In particular, an elementary quantum 
system is defined by a unitary irreducible representation of the Galilean line group. As stated above, if such a representation is to contain a unitary irreducible \emph{projective} representation of the Galilei 
group as a subrepresentation, then the relevant representation of the Galilean line group must have a sufficiently rich cocycle structure. 

It was stated in  \cite{sw1} that a two-cocycle of the Galilean line group that reduces to a non-trivial two-cocycle of the Galilei group appears to exist only when rotations are time independent. Consequently, the 
representations reported in \cite{sw1} were constructed under this restriction.  The main mathematical difficulty with time dependent rotations is that 
time dependent space translations $\bs{a}$ and their derivatives $\dot{\bs{a}}$ do not transform the same  way under rotations owing to the inhomogeneous term in the derivative
$\frac{d}{dt}R(t)\bs{a}(t)=R(t)\dot{\bs{a}}(t)+\dot{R}(t)\bs{a}(t)$.  On the other hand, we expect time dependent rotations to generate coriolis forces, but not fictitious forces that can be simulated by gravitational potentials. Since our focus on this paper is the fictitious forces and role of the equivalence principle, there is no compelling physical motivation to include time dependent rotations in the first place. Therefore, we will henceforth consider only linear accelerations. While all of our results remain valid for constant rotations, in order to avoid
irrelevant complications, in the following we will suppress rotations entirely. 

Setting $R=I$, the representations constructed in \cite{sw1} can be defined by the transformation 
formula for the generalized velocity eigenvectors $\mid\bs{q}\rangle$, 
\begin{equation}
\hat{U}^\times(\bs{a},b)\mid\bs{q}\rangle=e^{i\xi(g;\bs{q})}\mid\Lambda_{-b}\bs{q}'\rangle\label{2.9},
\end{equation}
where 
\begin{eqnarray}
\xi(g;\bs{q})&=&m(\bs{q}'\cdot\bs{a}-\frac{1}{2}\bs{a}\cdot{\dot{\bs{a}}}+\frac{1}{2}(\Lambda_{-b}-1)\bs{q}'\cdot{\bs{a}}_{\bs{q}'})-wb\nonumber\\
\bs{q}'&=&\bs{q}+\dot{\bs{a}},\label{2.9b}
\end{eqnarray}
or by the transformation formula for the $L^2$-functions $\psi(\bs{q})$, 
\begin{equation}
\Bigl(\hat{U}(\bs{a},b)\psi\Bigr)(\bs{q})=e^{-i\xi(g^{-1};\bs{q})}\psi(\Lambda_{b}\tilde{\bs{q}})\label{2.10},
\end{equation}
where
\begin{eqnarray}
\xi(g^{-1};\bs{q})&=&-\frac{1}{2}m\Bigl\{(\Lambda_b+1)\bs{q}\cdot\Lambda_{-b}\bs{a}-
\bs{a}\cdot{\dot{\bs{a}}}-(\Lambda_b-1)\bs{q}\cdot\bs{a}_{\bs{q}}\Bigr.\nonumber\\
&&\quad\Bigl.-(\Lambda_{-b}\dot{\bs{a}}\cdot\bs{a}_{\bs{q}}-\dot{\bs{a}}\cdot\Lambda_b\bs{a}_{\bs{q}})\Bigr\}+wb\nonumber\\
\tilde{\bs{q}}&=&\bs{q}-\Lambda_{-b}\dot{\bs{a}}.\label{2.10b}
\end{eqnarray}
In \eqref{2.9b} and \eqref{2.10b}, $\bs{a}_{\bs{q}}$ is the rotation-free Euclidean line group element that boosts the zero velocity vector to an arbitrary, time-dependent velocity  $\bs{q}$:
\begin{equation}
\dot{\bs{a}}_{\bs{q}}=\bs{q}\label{2.10c}
\end{equation}

When time translations are suppressed (so that, in particular, $\Lambda_b=1$), this representation reduces to:
\begin{equation}
\Bigl(\hat{U}(\bs{a})\psi\Bigr)(\bs{q})=e^{im\left(\bs{q}\cdot\bs{a}-\frac{1}{2}\bs{a}\cdot\dot{\bs{a}}\right)-iwb}\psi(\bs{q}-\dot{\bs{a}})\label{2.11}
\end{equation}
This is the representation constructed in \cite{bk1}, except for the trivial phase factor $e^{-i\frac{m}{2}\bs{a}\cdot\dot{\bs{a}}-iwb}$.  Further, when restricted to Galilean transformations, 
this representation reduces to a well-known projective representation of the Galilei group \cite{levy, bk2}.

The Hamiltonian $\hat{H}=i\left.\frac{\partial{\hat{U}(I,\bs{0},b)}}{\partial b}\right|_{b=0}$ for the representation \cite{sw1} is
 \begin{equation}
\left(\hat{H}\psi\right)(\bs{q})=\left(\frac{\hat{\bs{P}}^2}{2m}+\hat{V}+m\dot{\bs{q}}\cdot\left(\hat{\bs{X}}+\frac{1}{2}{\bs{a}_{\bs{q}}}\right)\right)\psi(\bs{q})\label{2.12}.
\end{equation}
where $\hat{\bs{X}}:=i\nabla_{\bs{p}}=i\frac{1}{m}\nabla_{\bs{q}}$ is the position operator canonically conjugated to the momentum operator. The third term  in the Hamiltonian 
is  the fictitious potential energy term that arises from the non-inertial character of the reference frames.  As expected, it is proportional to inertial mass $m$ and acceleration $\dot{\bs{q}}$.
Moreover, unlike in the Galilei case, here $[\hat{H},\hat{\bs{P}}]\not=0$. 

As mentioned above, these results and conclusions critically depend on the choice of two-cocycle made for the representation. In the next section, we will consider a more general class of 
two-cocycles and unitary representations of the Galilean line group associated with these.

\section{General representations of the Galilean line group}\label{sec4}
Our main goal in this paper is to study cocycle representations of the Galilean line group ${\cal G}$, specifically to examine 
if there exist representations more general than those constructed in \cite{bk1,sw1} and if these representations also uphold 
the equivalence principle. This problem is tantamount to determining 
the second cohomology group of ${\cal G}$ with respect to a suitable homomorphism and extension group $A$. See appendix A for a brief discussion of the general mathematical theory of group extensions, 
including the definitions of the terms cochains, coboundaries and cocycles that we will extensively use in this section. 

\subsection{Two-cocycles of the Galilean line group}\label{sec4.1}
Elements of  ${\cal G}$ include rotations $R$ and spatial translations $\bs{a}$ that are functions of time. Therefore, in general, 
we expect cochains of the line group to take values on a group that consists of functions of time. The simplest choice is the additive group of real, scalar functions of time, 
taken to be analytic for the  sake of simplicity. Let us denote this group by ${\cal F}(\mathbb{R})$ and a generic element thereof by $\varphi$. When the argument 
of $\varphi$ is not explicitly needed, we will  refer to the group simply by ${\cal F}$. 

In order to define 
the composition rule for the Galilean line group extended by ${\cal F}({\mathbb{R}})$, we must consider the relevant 
homomorphism from the Galilean line group to the automorphism group of ${\cal F}({\mathbb{R}})$.  Note that the composition rule  \eqref{2.5} for ${\cal G}$
involves the shift operator $\Lambda_b$, which simply accounts 
for the fact that a second group element gets evaluated at the point in time translated by the first group element, $t+b_1$. This property must also hold for the functions $\varphi$ of ${\cal F}(\mathbb{R})$--that 
is to say, the natural homomorphism  $\sigma:\ {\cal G}\to \text{Aut}\left({\cal F}({\mathbb{R}})\right)$ is simply the shift operator $\Lambda_b$. 
Thus, we consider extensions of ${\cal G}$ by ${\cal F}({\mathbb{R}})$, defined by the composition rule
\begin{equation}
(\varphi_2,g_2)(\varphi_1,g_1)=(\Lambda_{b_1}\varphi_2+\varphi_1+\omega(g_2,g_1), g_2g_1)\label{4.1}
\end{equation}
with $\omega\in H_\Lambda^2({\cal G}, {\cal F})$. It satisfies the two-cocycle condition 
\begin{equation}
\Lambda_{b_1}\omega(g_3,g_2)+\omega(g_3g_2,g_1)=\omega(g_2,g_1)+\omega(g_3,g_2g_1),\label{4.2}
\end{equation}

Further, any acceptable two-cocycle, in addition to satisfying \eqref{4.2}, must reduce to a two-cocycle of the Galilei group, such as \cite{levy}
\begin{equation}
\omega(g_2,g_1)=\frac{1}{2}m(\bs{a}_2\cdot\bs{v}_1-\bs{v}_2\cdot\bs{a}_1+b_1\bs{v}_2\bs{v}_1),\label{4.3}
\end{equation}
when $g_2$ and $g_1$ consist  of spatial translation functions of the form $\bs{a}(t)=\bs{a}+\bs{v}t$. For the reasons discussed above in section \ref{sec2}, we have suppressed rotations 
in \eqref{4.3} and in the discussion below. 

The cocycle chosen in \cite{sw1} that fulfills \eqref{4.2} and \eqref{4.3} is the following:
\begin{equation}
\omega(g_2, g_1) =\frac{1}{2}m\Bigl(( \Lambda_{b_1}\bs{a}_2)\cdot \dot{\bs{a}}_1 - (\Lambda_{b_1} \dot{\bs{a}}_2)\cdot  \bs{a}_1\Bigr),\label{4.4}. 
\end{equation}
If time translations are suppressed (so that $\Lambda_b=1$), \eqref{4.4} becomes the same as the cocycle chosen in \cite{bk1}, up to a two-coboundary.  This choice of cocycle 
was critical to the conclusion arrived at in \cite{bk1,sw1} that the equivalence principle can be consistent with the axioms of quantum mechanics. 

However, as we now show, there exist more general cocycles. To that end, let $\bs{B}(\bs{a})$ and $\bs{C}(\bs{a})$ be two vector valued functions of the acceleration transformations and define
\begin{equation}
\omega(g_2,g_1)=\frac{1}{2}\left(\Lambda_{b_1}\bs{B}(\bs{a}_2)\right)\cdot\bs{C}(\bs{a}_1)-\frac{1}{2}\left(\Lambda_{b_1}\bs{C}(\bs{a}_2)\right)\cdot\bs{B}(\bs{a}_1)\label{4.5}
\end{equation}
If we demand that this function fulfill the two-cocycle condition \eqref{4.2}, we obtain the following constraints on $\bs{B}(\bs{a})$ and $\bs{C}(\bs{a})$:
\begin{eqnarray}
\Lambda_{b}\bs{B}(\bs{a})&=&\bs{B}(\Lambda_{b}\bs{a})\nonumber\\
\Lambda_{b}\bs{C}(\bs{a})&=&\bs{C}(\Lambda_{b}\bs{a})\nonumber\\
\bs{B}(\bs{a}_2+\bs{a}_1)&=&\bs{B}(\bs{a}_2)+\bs{B}(\bs{a}_1)\nonumber\\
\bs{C}(\bs{a}_2+\bs{a}_1)&=&\bs{C}(\bs{a}_2)+\bs{B}(\bs{a}_1)\label{4.6}
\end{eqnarray}
The requirement that \eqref{4.5} have the correct reduction to \eqref{4.3} imposes further constraints. However, 
before we consider this embedding, let us construct unitary irreducible representations of the Galilean line group for the general cocycle \eqref{4.5}. 
\subsection{Unitary cocycle representations of the Galilean line group}\label{sec4.2}
The method of induced representations 
made use of in \cite{sw1} can be followed here, too, and the construction of the representation runs parallel to that of \cite{sw1}. Therefore, we omit a detailed 
derivation here and refer the reader to \cite{sw1}. As in \cite{bk1,sw1}, 
we restrict our discussion to a spin zero representation. 

The representation corresponding to \eqref{4.5}  is defined by the following 
transformation formula for the generalized velocity eigenvectors $\mid\bs{q}\rangle$:
\begin{equation}
\hat{U}^\times(\bs{a},b)\mid\bs{q}\rangle=e^{i\xi(g,\bs{q})}\mid\Lambda_{-b}(\bs{q}+\bs{C}(\bs{a}))\rangle\label{4.7}
\end{equation}
where
\begin{eqnarray}
\xi(g,\bs{q})&=&\bs{B}(\bs{a})\cdot(\bs{q}+\bs{C}(\bs{a}))-\frac{1}{2}\bs{B}(\bs{a})\cdot\bs{C}(\bs{a})\nonumber\\
&&\quad+\frac{1}{2}(\Lambda_{-b}-1)\bs{B}(\bs{a}_{\bs{q}'})\cdot\bs{C}(\bs{a}_{{\bs{q}'}})-wb\nonumber\\
\bs{q}'&=&\bs{q}+\bs{C}(\bs{a})\label{4.8}
\end{eqnarray}
Here, $w$ is the internal energy of the system so that at rest the time translations 
are represented by $U(b)\mid\bs{0}\rangle=e^{-iwb}\mid\bs{0}\rangle$. 
The spatial translation vector $\bs{a}_{\bs{q}}$ is defined as that which boosts the zero velocity vector 
to $\bs{q}$ by way of $\bs{C}$ (Cf.~\eqref{2.10c}): 
\begin{equation}
\bs{C}(\bs{a}_{\bs{q}})=\bs{q}\label{4.9}
\end{equation}
By using the duality formula 
\begin{equation}
\langle \hat{U}(g)\psi\mid\bs{q}\rangle=\langle\psi\mid\hat{U}^\times(g^{-1})\bs{q}\rangle\label{4.10}
\end{equation}
we can also obtain the representation in terms of the transformation properties of the velocity wave functions $\psi(\bs{q})=\langle\bs{q}\mid\psi\rangle$: 
\begin{eqnarray}
\left(\hat{U}(\bs{a},b))\psi\right)(\bs{q})&=&\langle\hat{U}^\times((\bs{a},b)^{-1})\bs{q}\mid\psi\rangle\nonumber\\
&=&e^{-i\xi(g^{-1},\bs{q})}\psi(\Lambda_b\bs{q}-\bs{C}(\bs{a}))\label{4.11}
\end{eqnarray}
where, from \eqref{4.8}, 
\begin{eqnarray}
\xi(g^{-1},\bs{q})&=&-\Lambda_{-b}\Bigl(\bs{B}(\bs{a})\cdot(\bs{q}-\bs{C}(\bs{a}))\Bigr)-\frac{1}{2}\Lambda_{-b}\Bigl(\bs{B}(\bs{a})\cdot\bs{C}(\bs{a})\Bigr)\nonumber\\
&&\quad+\frac{1}{2}(\Lambda_b-1)\bs{B}(\bs{a}_{\tilde{\bs{q}}})\cdot\bs{C}(\bs{a}_{\tilde{\bs{q}}})+wb\nonumber\\
\tilde{\bs{q}}&=&\bs{q}-\Lambda_{-b}\left(\bs{C}(\bs{a})\right)
\label{4.12}
\end{eqnarray}

\subsection{Hamiltonian, momentum and boost operators}\label{sec4.4}
We can now calculate the action of the generators on the representation space by differentiating each one-parameter operator subgroup with respect to that parameter 
and evaluating the derivative at the group identity. Thus, let us consider  \eqref{4.12} (or, \eqref{4.11}) for $g=(I,\bs{0},b)$:
\begin{equation}
\left(\hat{U}(b)\psi\right)(\bs{q})=e^{-\frac{i}{2}(\Lambda_b-1)\bs{B}(\bs{a}_{\bs{q}})\cdot\bs{q}}e^{-iwb}\psi(\Lambda_b\bs{q}),\label{4.13}
\end{equation}
where we have used the defining identity \eqref{4.9}. We then obtain the Hamiltonian by 
\begin{eqnarray}
\left(\hat{H}\psi\right)(\bs{q})&:=&\left(\left.i\frac{d\hat{U}(b)}{db}\right|_{b=0}\psi\right)(\bs{q})\nonumber\\
&=&\left(\frac{1}{2}\frac{d}{dt}\left(\bs{B}(\bs{a}_{\bs{q}})\cdot\bs{q}\right)+w+i\dot{\bs{q}}\cdot\nabla\right)\psi(\bs{q})\nonumber\\
&=&\left(\frac{1}{2}\frac{d\bs{B}(\bs{a}_{\bs{q}})}{dt}\cdot\bs{q}+w+\dot{\bs{q}}\cdot(i\nabla+\bs{B}(\bs{a}_{\bs{q}}))\right)\psi(\bs{q})
\label{4.14}
\end{eqnarray}
That the generator of the automorphism group $\Lambda_b$ is the derivative with respect to time, i.e., $\frac{d\Lambda_b}{db}=\Lambda_b\frac{d}{dt}=\frac{d}{dt}\Lambda_b$, 
has been used to obtain the second equality of \eqref{4.14}. 

Similarly, to obtain the momentum operator, we consider time independent space translations, $\bs{a}^{0}$. Then, with $g=(I,\bs{a}^0, 0)$, \eqref{4.11} gives
\begin{equation}
\left(\hat{U}(\bs{a}^0)\psi\right)(\bs{q})=e^{i\bs{B}(\bs{a}^0)\cdot(\bs{q}-\bs{C}(\bs{a}^0))+\frac{i}{2}\bs{B}(\bs{a}^0)\cdot\bs{C}(\bs{a}^0)}\psi(\bs{q}-\bs{C}(\bs{a}^0))\label{4.15}
\end{equation}
From this expression, we can obtain the momentum operator as usual: 
\begin{eqnarray}
\left(\hat{P}_i\psi\right)(\bs{q})&:=&\left(-\left.i\frac{d\hat{U}(\bs{a}^0)}{da^0_i}\right|_{\bs{a}^0=0}\psi\right)(\bs{q})\nonumber\\
&=&\left(\left.\frac{\partial\bs{B}(\bs{a}^0)}{\partial a^0_i}\right|_{\bs{a}^0=0}\cdot\bs{q}+i\left.\frac{\partial\bs{C}(\bs{a}^0)}{\partial a^0_i}\right|_{\bs{a}^0=0}\cdot\nabla\right)\psi(\bs{q})\label{4.16}
\end{eqnarray}

For the representations of the Galilei group, the boost operator $\hat{\bs{K}}$ is defined in the same vein by differentiating the group of velocity transformations $\hat{U}(\bs{v})$. When scaled by the mass, the boost 
operator acquires interpretation as position, $\hat{\bs{X}}=\frac{1}{m}\hat{\bs{K}}$. The position and momentum operators obtained this way fulfill the Heisenberg commutation relations. 
For the Galilean line group, on the other hand, there are infinitely many such boost operators, corresponding to each Taylor coefficient of time dependent spatial translations $\bs{a}(t)=\sum_{n=0}^\infty \frac{1}{n!}\bs{a}^{(n)}t^n$, where $\bs{a}^{(n)}=\left.\frac{d^n\bs{a}}{dt^n}\right|_{t=0}$.  Hence, for $n=1,2,3\cdots$, 
\begin{eqnarray}
\hat{K}^{(n)}_i&:=&\left(-\left.i\frac{d\hat{U}(\bs{a})}{da^{(n)}_i}\right|_{\bs{a}=0}\psi\right)(\bs{q})\nonumber\\
&=&\left(\left.\frac{\partial\bs{B}(\bs{a})}{\partial a^{(n)}_i}\right|_{\bs{a}=0}\cdot\bs{q}+i\left.\frac{\partial\bs{C}(\bs{a})}{\partial a^{(n)}_i}\right|_{\bs{a}=0}\cdot\nabla\right)\psi(\bs{q})\label{4.17}
\end{eqnarray}
Note that $\bs{P}$ is simply the $n=0$ case of this general expression. Before we consider the commutation relations among these operators, let us discuss the reduction of the representation to a projective 
representation of the Galilei group. 

\subsection{Reduction to the Galilean case and equivalence principle}\label{sec4.5}
Our analysis so far has been quite general in that we have not demanded anything beyond the two-cocycle condition \eqref{4.2} for the function \eqref{4.5}, or equivalently \eqref{4.6} for $\bs{B}(\bs{a})$ and 
$\bs{C}(\bs{a})$. However, while necessary, \eqref{4.2} is not a sufficient condition for the function \eqref{4.5} to lead to an extension of the Galilean line group because it is only non-trivial 
 two-cocycles (i.e., elements of the second cohomology group or  two-cocycles that are not two-coboundaries) that define extensions of a group. In addition, as stated above, an acceptable  two-cocycle \eqref{4.5} 
 must also have the property that it reduces to a non-trivial two-cocycle of the Galilei group when $\bs{a}(t)=\bs{a}_0+\bs{v}t$ for constant $\bs{a}_0$ and $\bs{v}$. But any two-cocycle \eqref{4.5} that reduces to a 
 non-trivial cocycle of the Galilei group is necessarily a non-trivial two-cocycle of the line group. Therefore, it only remains for us to establish that there exist functions \eqref{4.5} that reduce to \eqref{4.3}. 

To that end, recall  that the two-cocycle \eqref{4.4} used in \cite{sw1} has the required reduction property. The main structural feature of \eqref{4.4} that ensures the reduction to \eqref{4.3} is the appearance of the derivatives $\dot{\bs{a}}$. This suggests that $\bs{B}$ and $\bs{C}$ contain derivatives of $\bs{a}$. Therefore, as a special class, 
we consider $\bs{B}(\bs{a})$ and $\bs{C}(\bs{a})$ of the form
\begin{eqnarray}
\bs{B}(\bs{a})&=&\sum_{n=0}^\infty \beta_n\frac{d^n\bs{a}}{dt^n}\nonumber\\
\bs{C}(\bs{a})&=&\sum_{n=0}^\infty \gamma_n\frac{d^n\bs{a}}{dt^n}\label{4.18}
\end{eqnarray}
where $\beta_n$ and $\gamma_n$ are arbitrary constants. 

With $\bs{B}(\bs{a})$ and $\bs{C}(\bs{a})$ as defined by \eqref{4.18}, for the two-cocycle \eqref{4.5} to reduce to \eqref{4.3}, we must have the following constraint on the constants $\beta_0$, $\beta_1$, $\gamma_0$ and $\gamma_1$:
\begin{equation}
\beta_0\gamma_1-\gamma_0\beta_1=m\label{4.19}
\end{equation}
where $m$ is the inertial mass. There are no constraints on the remaining $\beta_n$ and $\gamma_n$. 

Furthermore, with \eqref{4.18}, the expression \eqref{4.14}  for the Hamiltonian becomes
\begin{eqnarray}
\left(\hat{H}\psi\right)(\bs{q})&=&\left(\frac{1}{2}\sum_{n=0}^\infty \beta_n\frac{d^{n+1}(\bs{a}_{\bs{q}})}{dt^{n+1}}\cdot\bs{q}+w+\dot{\bs{q}}\cdot(i\nabla+\sum_{n=0}^\infty \beta_n\frac{d^n \bs{a}_{\bs{q}}}{dt^n}\right)\psi(\bs{q})\nonumber\\
\label{4.20}
\end{eqnarray}
Similarly, substituting \eqref{4.18}  in \eqref{4.16} and \eqref{4.17}, we obtain,
\begin{eqnarray}
\left(\hat{\bs{P}}\psi\right)(\bs{q})&=&(\beta_0\bs{q}+\gamma_0i\nabla)\psi(\bs{q})\nonumber\\
\left(\hat{\bs{K}}^{(n)}\psi\right)(\bs{q})&=&\sum_{k=0}^n\frac{1}{(n-k)!}\left(\beta_k t^{n-k} \bs{q}+i\gamma_k t^{n-k}\nabla\right)\psi(\bs{q})\label{4.21}
\end{eqnarray}
Using these expressions, we can compute the commutation relations between momentum and Galilean boost operators: 
\begin{eqnarray}
\left[\hat{K}^{(1)}_i, \hat{P}_j\right]&=&i(\beta_0\gamma_1-\beta_1\gamma_0)\delta_{ij}\nonumber\\
&=&im\delta_{ij}\label{4.22}
\end{eqnarray}
where the last equality follows from \eqref{4.19}, the condition that ensures that our general cocycle representation has the proper reduction to the Galilean case with mass $m$. 

These expressions show that for the class of general representations of the Galilean line group we have considered, the Heisenberg commutation relations are always fulfilled. 
This is clearly a direct consequence of the requirement that these general representations contain a representation of the Galilei group, the condition that allows us to define 
inertial mass and its relationship \eqref{4.19} to the two-cocycle. However, even with this embedding, the equivalence principle is not upheld in these representations. 
In particular, \eqref{4.18} shows that the Hamiltonian does not decompose into the usual sum of kinetic energy, potential energy and a fictitious potential energy that is proportional 
to inertial mass. The presence of the infinite family of constants $\beta_n$ shows that the contribution to the Hamiltonian that results from the non-inertial effects of
 the reference frame need not have a single coupling constant, let alone that it be the inertial mass. 

We may make a choice, however, for $\beta_0$, $\gamma_0$ and $\gamma_1$ under the constraint \eqref{4.19} such that the expression for the momentum operator is simplified: 
\begin{equation}
\beta_0=m,\ \gamma_1=1,\ \gamma_0=0\label{4.23}
\end{equation}
Then, 
 \begin{eqnarray}
 \bs{B}(\bs{a})&=&m\bs{a}+\sum_{n=1}^\infty \beta_n\frac{d^n\bs{a}}{dt^n}\nonumber\\
\bs{C}(\bs{a})&=&\frac{d\bs{a}}{dt}+\sum_{n=2}^\infty \gamma_n\frac{d^n\bs{a}}{dt^n}\label{4.24}
\end{eqnarray} 
and the momentum operator acquires the form
\begin{equation}
\left(\hat{\bs{P}}\psi\right)(\bs{q})=m\bs{q}\psi(\bs{q})\equiv\bs{p}\psi(\bs{q})\label{4.25}
\end{equation}
This expression, along with the interpretation that $m$ is the inertial mass, justifies the interpretation of $\bs{q}$ as the (time dependent) velocity. Under \eqref{4.23}, the Galilean boost operator 
has the form 
\begin{equation}
\hat{\bs{K}}^{(1)}=\left(t+\frac{\beta_1}{m}\right)\hat{\bs{P}}+i\nabla_{\bs{q}}=\left(t+\frac{\beta_1}{m}\right)\hat{\bs{P}}+m\hat{\bs{X}}\label{4.26}
\end{equation}
where $\hat{\bs{X}}=i\nabla_{\bs{p}}=\frac{i}{m}\nabla_{\bs{q}}$. In terms of these new operators, we may rewrite the Hamiltonian as 
\begin{eqnarray}
\left(\hat{H}\psi\right)(\bs{q})&=&\left(\frac{\hat{\bs{P}}^2}{2m}+\hat{V}+m\dot{\bs{q}}\cdot\left(\hat{\bs{X}}+\bs{a}_{\bs{q}}\hat{I}+\sum_{n=1}^\infty \frac{\beta_n}{m}\frac{d^{n}(\bs{a}_{\bs{q}})}{dt^{n}}\right)\right.\nonumber\\
&&\left.+\frac{1}{2}\hat{\bs{P}}\cdot\left(\frac{1}{m}\sum_{n=1}^\infty\beta_n\frac{d^n\bs{a}_{\bs{q}}}{dt^n}-\sum_{n=2}^\infty\gamma_n\frac{d^n\bs{a}_{\bs{q}}}{dt^n}\right)\right)\psi(\bs{q})\label{4.27}
\end{eqnarray}
where we have introduced the internal energy operator $\hat{V}\psi(\bs{q})=w\psi(\bs{q})$, which, by construction, is an invariant of the Galilean line group.  The last two terms clearly show that the equivalence principle is not upheld in the general representations of the Galilean line group. If we set $\beta_n=0$ for $n=1,2,3,\cdots$ and $\gamma_n=0$ for $n=2,3,4,\cdots$, then the general representations constructed here reduce to those constructed in \cite{sw1}.  The Hamiltonian \eqref{4.27} also reduces to \eqref{2.12} and the equivalence principle is restored. 

\subsection{Time evolution of expectation values}\label{sec4.6}
While the ``fictitious potential energy terms" of the Hamiltonian lead to violations of the equivalence principle, these violations disappear from the equations of motion for the expectation values of position and momentum operators. Straightforward calculations using the general dynamical equation $\frac{d}{db}\langle\psi\mid\hat{A}\mid\psi\rangle=i\langle\psi\mid[\hat{H},\hat{A}]\psi\rangle$ and the Hamiltonian \eqref{4.27} give:
\begin{eqnarray}
\frac{d}{db}\langle\hat{\bs{P}}\rangle&=&m\dot{\bs{q}}\label{4.28}\\
\frac{d}{db}\langle\hat{\bs{X}}\rangle&=&\frac{1}{m}\langle\hat{\bs{P}}\rangle+\frac{1}{2}\left(\sum_{n=1}^\infty\frac{\beta_n}{m}\frac{d^{n+1}\bs{a}_{\bs{q}}}{dt^{n+1}}-\sum_{n=2}^\infty\gamma_n\frac{d^n{\bs{a}_{\bs{q}}}}{dt^n}\right)\label{4.29}\\
\frac{d^2}{db^2}\langle\hat{\bs{X}}\rangle&=&\frac{d}{db}\frac{\langle{\hat{\bs{P}}}\rangle}{m}=\dot{\bs{q}}\label{4.30}
\end{eqnarray}
The last equation shows that we do not detect any violations of the equivalence principle in the acceleration of the expectation value of position.

\section{The equivalence principle in classical mechanics}\label{sec5}
In this section, we examine the classical Hamiltonian mechanics counterpart of the previous results. As is well known,  Hamilton's equations 
of motion are form invariant under all canonical transformations of generalized coordinates and conjugate momenta, a much larger group of transformations 
than that of Galilean transformations. The group of canonical transformations includes the acceleration transformations of the type we considered above. 

Let us suppose that $\bs{x}$ are the Cartesian coordinates of some inertial reference frame and let $\bs{p}$  be the momenta canonically conjugated to $\bs{x}$. Likewise, let 
$\bs{x}'$ and $\bs{p}'$ be the canonical coordinates and momenta in a reference frame linearly accelerating with respect to the first. In order to obtain the 
canonical transformations between $(\bs{x},\bs{p})$ and $(\bs{x}',\bs{p}')$, consider  a generating function of the form 
\begin{equation}
F(\bs{x},\bs{p}',t)=\left(\bs{x}+\bs{B}(\bs{a}(t))\right)\bs{p}'+g\left(\bs{x};\bs{a}(t),\dot{\bs{a}}(t),\ddot{\bs{a}}(t),\cdots\right)\label{5.1}
\end{equation}
where $g$ is an arbitrary smooth function of $\bs{x}$, time dependent spatial translations $\bs{a}(t)$ and all of their derivatives $\frac{d^n\bs{a}(t)}{dt^n}$. For notational convenience, let us denote the argument of $g$ more generally 
as $g(\bs{x},\bs{z})$, where $\bs{z}=(\bs{z}_1,\bs{z}_2,\bs{z}_3,\cdots)$. To obtain the generating function  \eqref{5.1}, we set $\left.g(\bs{x},\bs{z})\right|_{\bs{z}_n=\frac{d^n\bs{a}}{dt^n}}$, the evaluation of $g$ at $\bs{z}_n=\frac{d^n\bs{a}}{dt^n}$.

The generating function \eqref{5.1} defines canonical transformations between $(\bs{x},\bs{p})$ and $(\bs{x}',\bs{p}')$ when $\bs{B}(\bs{a})$ is an arbitrary function of $\bs{a}(t)$ and its derivatives. As seen from \eqref{5.2} below, if we wish to recover  the coordinate transformations \eqref{2.3} (with $R(t)=I$) that defined linear acceleration transformations, then we must clearly choose $\bs{a}(t)$ for $\bs{B}(\bs{a})$. However, we take  it to be the function defined by \eqref{4.18} because this choice makes the boost operators \eqref{4.21} and their corresponding classical generators have the same form (see \eqref{5.12} below). The primitive coordinate transformations \eqref{2.3} can be recovered by setting $\beta_n=\delta_{n0}$.

  The canonical transformations are given by the following well-known expressions: 
  \begin{eqnarray}
 x'_i&=&\frac{\partial F}{\partial p'_i}\nonumber\\
 &=&x_i+B_i(\bs{a})\label{5.2}\\
 p_i&=&\frac{\partial F}{\partial x_i}\nonumber\\
 &=&p'_i+\frac{\partial g}{\partial x_i}\label{5.3}\\
 H'(\bs{x}',\bs{p}')&=&H(\bs{x},\bs{p})+\frac{\partial F(\bs{x},\bs{p}')}{\partial t}\nonumber\\
 &=&H(\bs{x},\bs{p})+\frac{\partial}{\partial t}\bs{B}(\bs{a})\cdot\bs{p}'+\frac{\partial g}{\partial t}\label{5.4}.
 \end{eqnarray}
It is understood that $\bs{x}$ and $\bs{p}$ on the right hand side of \eqref{5.4} are to be written in terms of  $\bs{x}'$ and $\bs{p}'$  before the dynamics of a physical system can be discussed in terms of these transformed dynamical variables. The expressions for $\bs{x}$ and $\bs{p}$ as functions of $\bs{x}'$ and $\bs{p}'$ are to be obtained from \eqref{5.2} and \eqref{5.3}. Thus, 
\begin{eqnarray}
 H'(\bs{x}',\bs{p}')&=&H\left(\bs{x}'-\bs{B}(\bs{a}), \bs{p}'+\left.\nabla_{\bs{x}}g(\bs{x},\bs{z})\right|_{\bs{z}_n=\frac{d^n\bs{a}}{dt^n},\ \bs{x}=\bs{x}'-\bs{B}(\bs{a})}\right)+\frac{\partial}{\partial t}\bs{B}(\bs{a})\cdot\bs{p}'\nonumber\\
 &&\quad+\sum_{n=0}^\infty \left.\frac{d^{n+1}\bs{a}(t)}{dt^{n+1}}\cdot\nabla_{\bs{z}_n}g(\bs{x},\bs{z})\right|_{\bs{z}_k=\frac{d^k\bs{a}}{dt^k},\ \bs{x}=\bs{x}'-\bs{B}(\bs{a})}\label{5.5}
 \end{eqnarray}

The equations of motion for the transformed dynamical variables $\bs{x}'$ and $\bs{p}'$ can be computed from Hamilton's equations, using the transformed Hamiltonian \eqref{5.5}. 
For the sake of definiteness, let us consider a free particle so the inertial frame Hamiltonian is $H(\bs{x},\bs{p})=\frac{1}{2m}\bs{p}^2$.  Then, 
\begin{eqnarray}
\dot{x}'_i&=&\frac{\partial H'}{\partial p'_i}\nonumber\\
&=&\frac{1}{m}\left(p'_i+\frac{\partial}{\partial x_i}g\right)+\frac{d}{dt}B_i(\bs{a})\label{5.6}\\
\dot{p'}_i&=&-\frac{\partial H'}{\partial x'_i}\nonumber\\
&=&-\frac{1}{m}\left(\bs{p}'+\nabla_{\bs{x}}g\right)\cdot\frac{\partial}{\partial x'_i}\nabla_{\bs{x}}g-\frac{\partial^2}{\partial x_i'\partial t}g\nonumber\\
&=&-\left(\dot{\bs{x}'}-\frac{d}{dt}\bs{B}(\bs{a})\right)\cdot\frac{\partial}{\partial x'_i}\nabla_{\bs{x}}g-\frac{\partial^2}{\partial x_i'\partial t}g\label{5.7}
\end{eqnarray} 
For simplicity, we have omitted the arguments of $g$ in these expressions. Taking the time derivative of \eqref{5.6}, 
\begin{equation}
\ddot{x}'_i=\frac{1}{m}\left(\dot{p}'_i+\left(\dot{\bs{x}'}-\dot{\bs{B}}(\bs{a}(t))\right)\cdot\nabla_{\bs{x}}\frac{\partial g}{\partial x_i}+\frac{\partial^2g}{\partial t\partial x_i}\right)+\ddot{\bs{B}}(\bs{a})\nonumber\\
\end{equation}
and substituting from \eqref{5.7}, we obtain
\begin{equation}
\ddot{\bs{x}'}=\ddot{\bs{B}}(\bs{a}(t))\label{5.8}
\end{equation}
This is clearly consistent with \eqref{5.2} and the equation of motion for a free particle in an inertial reference frame, $\ddot{\bs{x}}=0$. Further, if we set $\beta_n=\delta_{n0}$, then 
\begin{equation}
\ddot{\bs{x}'}=\ddot{\bs{a}}(t),
\end{equation}
the expression that we anticipate from the original definition of the acceleration transformations \eqref{2.3}. 
Therewith, we see that a generating function of the form \eqref{5.1} with an arbitrary function $g$ always gives of the same equation of motion for the position, regardless of the mass 
of the particle, and always satisfies the equivalence  principle. 

In order to examine the classical counterpart of the commutation relations calculated in section \ref{sec4}, let us treat Taylor coefficients of $\bs{a}$ as transformation parameters 
and calculate the corresponding generators (sometimes called the generators of infinitesimal canonical transformations). 
Further, let us choose $g$ to be of the form $g(\bs{x},\bs{a},\dot{\bs{a}},\ddot{\bs{a}},\cdots)=\bs{x}\cdot\bs{f}(\bs{a},\dot{\bs{a}},\ddot{\bs{a}},\cdots)$ for some arbitrary 
function $\bs{f}$ of $\bs{a}$ and all its derivatives. This choice will make Poisson brackets independent of coordinates and momenta. Let us also choose $\bs{B}(\bs{a})$ to be the function defined in \eqref{4.18}. Then, the generating function \eqref{5.1} has the form
\begin{eqnarray}
F(\bs{x},\bs{p}, t)&=&\bs{x}\cdot\bs{p}'+\bs{p}'\cdot\sum_{n=0}^\infty \beta_n\frac{d^n\bs{a}}{dt^n}+\bs{x}\cdot\bs{f}(\bs{a},\dot{\bs{a}},\ddot{\bs{a}},\cdots)\nonumber\\
&=&\bs{x}\cdot\bs{p}'+\bs{p}'\cdot\sum_{n=0}^\infty \bs{a}^{(n)}\sum_{k=0}^n \frac{\beta_kt^{n-k}}{(n-k)!}\nonumber\\
&&\quad +\left.\bs{x}\cdot\sum_{n=0}^\infty\sum_{k=0}^n (\bs{a}^{(n)}\cdot\nabla_{\bs{z}_k})\bs{f}(\bs{z})\right|_{\bs{z}_k=\frac{d^k\bs{a}}{dt^k}=0}\frac{t^{n-k}}{(n-k)!}\label{5.9}
\end{eqnarray}
where $\bs{a}^{(n)}$ are the Taylor coefficients of the function $\bs{a}(t)$. Corresponding to each $\bs{a}^{(n)}$, there is a generator $\bs{A}^{(n)}(\bs{x},\bs{p})$:
\begin{equation}
{A}_i^{(n)}(\bs{x},\bs{p})=\left.\sum_{j=1}^3x_j\sum_{k=0}^n\frac{\partial f_j}{\partial {{(z_k)_i}}}\right|_{\bs{z}_k=0}\frac{t^{n-k}}{(n-k)!}+p_i\sum_{k=0}^n\frac{\beta_kt^{n-k}}{(n-k)!}\label{5.10}
\end{equation}
The structure of these generators can be brought to a form quite similar to that of quantum boost generators \eqref{4.21} by choosing $\bs{f}$ such that 
\begin{equation}
\left.\frac{\partial f_j}{\partial {(z_k)}_i}\right|_{\bs{z}=0}=\delta_{ij}\gamma_k\label{5.11}
\end{equation}
This choice essentially sets $\bs{f}(\bs{a},\dot{\bs{a}},\ddot{\bs{a}}, \cdot)$ to be the same as $\bs{C}(\bs{a})$ of \eqref{4.18}. Then, \eqref{5.11} becomes 
\begin{equation}
\bs{A}^{(n)}(\bs{x},\bs{p})=\bs{x}\sum_{k=0}^n\frac{\gamma_kt^{n-k}}{(n-k)!}+\bs{p}\sum_{k=0}^n\frac{\beta_kt^{n-k}}{(n-k)!}\label{5.12}
\end{equation}
and the classical boost generators have the same form as the generators in the quantum case,  given by \eqref{4.21}. In particular, for the generator of pure space translations, we have 
\begin{equation}
\bs{A}^{(0)}(\bs{x},\bs{p})=\gamma_0\bs{x}+\beta_0\bs{p}\label{5.13}
\end{equation}
and for the generator of Galilei boosts,
\begin{equation}
\bs{A}^{(1)}(\bs{x},\bs{p})=(\gamma_1+\gamma_0t)\bs{x}+(\beta_1+\beta_0t)\bs{p}\label{5.14}
\end{equation}
Hence, we obtain the Poisson bracket
\begin{equation}
\left\{A^{(1)}_i,A^{(0)}_j\right\}=\beta_0\gamma_1-\beta_1\gamma_0\label{5.15}
\end{equation}
This expression is identical in form to \eqref{4.22}. If we impose here the constraint \eqref{4.19} that gives the Heisenberg commutation relations in the quantum case, 
we obtain the canonical Poisson brackets 
\begin{equation}
\left\{\frac{1}{m}A^{(1)}_i,A_j^{(0)}\right\}=\delta_{ij}\label{5.16}
\end{equation}

Note that in both classical and quantum mechanical Hamiltonians, there appears a term linear in momentum. Nevertheless, the equivalence principle is always upheld in the classical case in the sense that all trajectories are independent of the inertial mass and are identical to the trajectories that result from motion under an equivalent gravitational field. In the quantum case, while the inertial mass does not disappear from the state vector or wave function when it evolves under the Schroedinger equation, it is still meaningful to ask if the time evolution of the state vector governed by a Hamiltonian in a non-inertial reference frame is equivalent the time evolution in a gravitational field. To the extent that gravitational and inertial masses are equal, the evolution governed by a Hamiltonian such as \eqref{2.12} would be indistinguishable from the evolution in a suitable external gravitational potential. In contrast, such would clearly not be the case for the Hamiltonian \eqref{4.27} that follows from our more general representations, the equality of inertial and gravitational masses notwithstanding. 

\section{Concluding remarks}\label{sec6} 
One of the ways of grounding quantum theory is through the representations of relativity groups, groups that link different inertial reference frames.  As is well-known the relevant group for non-relativistic quantum theory is the Galilei group, discussed in section \ref{sec2}, valid when velocities are small compared to the speed of light and the curvature of the universe is not significant.  The irreducible projective representations of the Galilei group provide the operators that make up the basic structure of non-relativistic quantum theory, as well as supplying the Hilbert spaces on which these operators act.

But beyond the transformations linking different inertial reference frames are transformations to accelerating reference frames, leading to a group larger than the Galilei group which we call the Galilean line group.  This group also has irreducible two-cocycle representations, the unitary operators of which implement the acceleration transformations.  An important restriction on such representations is that they should act on the same representation spaces as the projective representations of the Galilei group (which are the physical one-particle Hilbert spaces);  further, when restricted to elements of the Galilei group, the resulting representation should be equivalent to the projective representation of the Galilei group.

Such a restriction however still leaves many possibilities for the Galilean line group.  In this paper we have exhibited a class of such representations and investigated the nature of the resulting fictitious forces.  In particular we have shown that in general the fictitious forces are not equivalent to gravitational potentials.  Such a violation of the non-relativistic equivalence principle seems all the more surprising in that there is no classical counterpart.  Specifically, we have shown that generating functions for acceleration transformations of linearly accelerated systems always lead to equations of motions that 
respect the equivalence principle.

One of the shortcomings of our analysis is that we have not included time dependent rotations.  It seems there are no two-cocycle representations of the Galilean line group that include rotational acceleration.  However, there do exist general representations of the Galilean line group that allow for time dependent rotations although they do not contribute to the two-cocycle, a result we will present in a subsequent publication.  Besides,  given the nature of rotational fictitious forces, there does not seem to be a natural relationship between time dependent rotations and gravitational potentials, the main reason for our focus 
 only on linear accelerations in this paper.

Of perhaps more interest is the nature of fictitious forces and their relationship to relativistic quantum theory, in which the Galilei group is replaced by the Poincare group.  Here also we have found interesting one-cocycle representations, a result which also will be presented in a subsequent publication.

\appendix
\section{Appendix: Cohomology and group extensions}\label{sec3}
Ever since the analysis of Bargmann, it is known that a projective representation of the Galilei group is equivalent to 
a true (vector) representation of a suitable central extension of the group. In this sense, it is the centrally extended Galilei group, rather than the group itself, that is of importance 
in quantum mechanics. Therefore, the requirement that the physically meaningful representations of the Galilean line group  contain a unitary projective representation of the Galilei group 
is tantamount to the requirement that we construct a suitable extension of the Galilean line group so as to  contain a given central extension 
of the Galilei group. As shown in \cite{sw1}, there exist no \emph{central} extensions of the Galilean line group that contain a given central extension of the Galilei group. However, as also shown in 
\cite{sw1},  there does exist a \emph{non-central} extension that contains a given central extension of the Galilei group. The non-centrality of this extension is due to the non-trivial homomorphism $(R,\bs{a},b)\to\Lambda_b$ from the  Galilean line group to the group of (analytic) scalar functions of time, the Abelian group which extends the line group. \subsection{Group extensions}\label{sec3.1} 
 The general problem of extending a group $G$ by another group $A$ can be stated as follows: find a group $\tilde{G}$ such that $A$ is an invariant su
 bgroup of $\tilde{G}$ and the quotient 
 $\frac{\tilde{G}}{A}$ is isomorphic to $G$. We may think of $\tilde{G}$ as the principal fiber bundle with structure group $A$. Then, if we choose a normalized trivializing section $s: G\to\tilde{G}$, then an element $\tilde{g}$ of 
 $\tilde{G}$ may be written as $\tilde{g}=as(g)\equiv (a,g)$ for some $a\in A$ and $g\in G$. The composition rule for $\tilde{G}$ reads
 \begin{eqnarray}
 \tilde{g}_2\cdot\tilde{g}_1&=&a_2s(g_2)a_1s(g_1)\nonumber\\
 &=&a_2(\sigma(g_2)a_1)\omega(g_2,g_1)s(g_2g_1)\nonumber\\
&\equiv&(a_2,g_2)(a_1,g_1)\nonumber\\
&=&(a_2(\sigma(g_2)a_1)\omega(g_2,g_1),g_2g_1)\label{3.1}
 \end{eqnarray}
where $\sigma(g)a=s(g)as(g)^{-1}$ is the automorphism of $A$ defined by $s(g)\in\tilde{G}$ and $\omega(g_2,g_1)$ is a unique element of $A$. 
The associativity of the product rule \eqref{3.1} under which $\tilde{G}$ is a group imposes the following 
restriction on  $\omega: \ G\times G\to A$: 
\begin{equation}
\omega(g_3,g_2)\omega(g_3g_2,g_1)=\left(\sigma(g_3)\omega(g_2,g_1)\right)\omega(g_3,g_2g_1)\label{3.2}
\end{equation}
A function $\omega: \ G\times G\to A$ satisfying \eqref{3.2} is called a factor system relative to (normalized) trivializing section $s$.  What is of interest to us in this paper involves the case where 
$A$ is an Abelian group. Then, writing \eqref{3.2} additively, we have 
\begin{equation}
\omega(g_3,g_2)+\omega(g_3g_2,g_1)=\sigma(g_3)\omega(g_2,g_1)+\omega(g_3,g_2g_1)\label{3.3}
\end{equation}
and therewith the composition rule \eqref{3.1} for $\tilde{G}$, 
\begin{equation}
(a_2,g_2)(a_1,g_1)=(a_2+\sigma(g_2)a_1+\omega(g_2,g_1), g_2g_1)\label{3.3b}
\end{equation}
For a given automorphism $\sigma(g)$, all inequivalent extensions of  $G$ by $A$ are determined by inequivalent mappings $\omega:\ G\times G\to A$. (Note that the discrepancy in the placement of the automorphism between  \eqref{4.1} and \eqref{3.3b} and between \eqref{4.2} and \eqref{3.3} is due to the composition structure of ${\cal G}$. See footnote \ref{footnote1}.) 

\subsection{The second cohomology group}\label{sec3.2}
The collection of inequivalent mappings that fulfill \eqref{3.3} constitute the second cohomology group $H_{\sigma}^2(G,A)$. In order to define it, we must first introduce the key notions of cochains, cocycles and coboundaries.Throughout this subsection, we let $A$ be an Abelian group and $G$, an arbitrary group. Our focus will be the algebraic structure of cohomology groups and we will not consider topological aspects.\\
{\bf n-cochains:} The mapping  $\alpha_n:\ G\times G\times\cdots\times G\to A$, i.e., 
\begin{equation}
\alpha_n: (g_1,g_2,\cdots, g_n)\to \alpha_n(g_1,g_2,\cdots, g_n)\in A\label{3.4}
\end{equation}
 is called an $n$-cochain. The collection of $n$-cochains form an Abelian group that we denote by $C^n(G,A)$, its composition rule defined by the pointwise composition of $\alpha_n(g_1,g_2,\cdots, g_n)$ as elements $A$. \\
{\bf Coboundary operator:} Let $\sigma:\ G\to \text{Aut}\ A$ be a homomorphism from $G$ to the automorphism group on $A$. 
The operator  $\delta_n:\ C^n(G,A)\to C^{n+1}(G,A)$  defined by the mapping (for the left action)
\begin{eqnarray}
(\delta_n\alpha_n)(g_1,g_2,\cdots,g_n,g_{n+1})&:=&\sigma(g_1)\alpha_n(g_2,g_3,\cdots,g_{n+1})\nonumber\\
&&\quad+\sum_{i=1}^n(-1)^i\alpha_n(g_1,\cdots,g_{i-1},g_i\cdot g_{i+1}, g_{i+2},\cdots, g_{n+1})\nonumber\\
&&\qquad +(-1)^{n+1}\alpha_n(g_1,g_2,\cdots,g_n)\label{3.5}
\end{eqnarray}
is called the coboundary operator. It is straightforward, albeit tedious, to show that 
\begin{equation}
\delta_{n+1}\circ\delta_n=0\label{3.6}
\end{equation}
Hence, 
\begin{equation}
\text{range}\ \delta_n\subset\ \text{kernel}\ \delta_{n+1}\label{3.7}
\end{equation}
{\bf Cocycles and coboundaries:} The inclusion \eqref{3.7} motivates the following definitions:
\begin{eqnarray}
\text{$n$-cocycles:}&&Z_\sigma^n(G,A):=\text{kernel}\ \delta_n\nonumber\\
\text{$n$-coboundaries:}&& B_\sigma^n(G,A):=\text{range}\ \delta_{n-1}\label{3.8}
\end{eqnarray}
Both $Z_\sigma^n(G,A)$ and $B_\sigma^n(G,A)$ are subgroups of $C^n(G,A)$. Further, by \eqref{3.7}, every $n$-coboundary is trivially an $n$-cocycle and we so have the inclusion 
$B_\sigma^n(G,A)\subset Z_\sigma^n(G,A)$. Since $A$ is Abelian, $B_\sigma^n(G,A)$ is in fact  an invariant subgroup of $Z_\sigma^n(G,A)$.\\
{\bf The $\bs{n^{\rm th}}$ cohomology group:} The quotient group
\begin{equation}
H_\sigma^n(G,A)=\frac{Z^n_\sigma(G,A)}{B_\sigma^n(G,A)}\label{3.9}
\end{equation}
is called the $n^{\rm th}$-cohomology group. An element of $H_\sigma^n(G,A)$ is a class of $n$-cocycles, equivalent up to $n$-coboundaries.\\
{\bf The second cohomology group:} A comparison of  \eqref{3.3} with the general formula \eqref{3.5} and the cocycle condition \eqref{3.8} shows 
that the constraint on the factor system that defines an extension of $G$ by $A$ is precisely the two-cocycle condition:
\begin{equation}
(\delta_2\alpha_2)(g_3,g_2,g_1):=\sigma(g_3)\alpha_2(g_2,g_1)-\alpha_2(g_3g_2,g_1)+\alpha_2(g_3,g_2g_1)-\alpha_2(g_3,g_2)=0\label{3.10}
\end{equation}
Moreover, from \eqref{3.3b} and \eqref{3.10}, it is possible to show that two cocycles $\alpha_2$ and $\alpha'_2$ that differ by a two-coundary, 
\begin{eqnarray}
\alpha_2(g_2,g_1)-\alpha'_2(g_2,g_1)&=&(\delta_1\alpha_1)(g_2,g_1)\nonumber\\
&=&\sigma(g_2)\alpha_1(g_1)-\alpha_1(g_2g_1)+\alpha_1(g_2),\label{3.11}
\end{eqnarray}
lead to equivalent extensions of $G$ by $A$. This means that, for a given $\sigma:\ G\to\text{Aut}\ A$, inequivalent extensions of $G$ by $A$ are in one-to-one correspondence with 
the elements of the second cohomology group $H_\sigma^2(G,A)$.   Note that while $A$ is an invariant subgroup of $\tilde{G}$, in general, $G$ is \emph{not} a subgroup of 
$\tilde{G}$. 

There are three special cases commonly encountered in the physics literature: 
\begin{enumerate}
\item {\bf Trivial $\sigma$ and non-trivial $\omega\in H^2(G,A)$:} In this case, the group law \eqref{3.3} becomes
\begin{equation}
(a_2,g_2)(a_1,g_1)=(a_2+a_1+\omega(g_2,g_1), g_2g_1)\label{3.12}
\end{equation}
In this case,  $\tilde{G}$ is called a \emph{central extension} of $G$ by $A$ and $A$ is a central subgroup of $\tilde{G}$. In particular, 
an extension of $G$ by $\mathbb{R}$ is central. 
\item {\bf Non-trivial $\sigma$ and trivial $\omega=0\in H_\sigma^2(G,A)$:} In this case, \eqref{3.3} becomes 
\begin{equation}
(a_2,g_2)(a_1,g_1)=(a_2+\sigma(g_2)a_1, g_2g_1)\label{3.13}
\end{equation}
This is the semidirect product of $G$ by $A$. In this case, $G$ is also a subgroup of $\tilde{G}$. 
\item {\bf Trivial $\sigma$ and trivial $\omega$:} The group law \eqref{3.3} now reads
\begin{equation}
(a_2,g_2)(a_1,g_1)=(a_2+a_1,g_2g_1)\label{3.14}
\end{equation}
This is the direct product of $G$ and $A$, both being invariant subgroups of $\tilde{G}$. 
\end{enumerate}

This summary provides the general mathematical context for the cocycle representations 
of the Galilean line group studied in section \ref{sec4}.  For more details, see references \cite{em1} and \cite{em2}. 
\newpage

\end{document}